\documentclass[12pt]{article}
\usepackage{amsmath}
\usepackage{amssymb}
\usepackage{amsthm}
\usepackage{amsbsy}
\usepackage{amsfonts}
\newtheorem{theorem}{Theorem}

\newtheorem{remark}[theorem]{Remark}

\newtheorem{example}[theorem]{Example}

\begin{document}

\title{On construction of discrete Darboux integrable  equations}
\author{ Kostyantyn Zheltukhin$^1$ and Natalya Zheltukhina$^2$\\
\small $^1$ Department of Mathematics, Faculty of Science,\\
\small Middle East Technical University,  Ankara, Turkey.\\
\small e-mail: zheltukh@metu.edu.tr\\
\small $^2$  Department of Mathematics, Faculty of Science,\\
\small Bilkent University, Ankara, Turkey.\\
\small e-mail: natalya@fen.bilkent.edu.tr}

\begin{titlepage}

\maketitle

\begin{abstract}
The problem of  discretization of Darboux integrable equations is considered.  Given a Darboux integrable continuous equation, one 
can obtain a Darboux  integrable differential-discrete equation, using the integrals of the continuous equation. In the present paper,
the discretization of the  differential-discrete equations is done using the corresponding characteristic algebras. 
New examples of integrable discrete equations are obtained.
\end{abstract}


\medskip


\end{titlepage}

\section{Introduction}

The problem of discretization of   the Darboux  integrable  hyperbolic equations  has generated a lot of interest in recent years.  
A hyperbolic equation
\begin{equation}\label{contEq}
w_{xy}=h(w,w_x,w_y)
\end{equation}
is called Darboux integrable if it admits two non-trivial functions 
$$
J(w,w_y,w_{yy},\dots)\quad \mbox{and} \quad G(w,w_x,w_{xx},\dots),
$$ depending on a finite number of variables,  
such that for all solutions of   (\ref{contEq}) we have
\begin{equation}
D_xJ=0 \quad \mbox{and} \quad D_yG=0,
\end{equation}
where  $D_x$ is the total $x-$derivative operator and  $D_y$ is the total $y-$derivative operator. The functions  $J(w,w_y,w_{yy},\dots)$  and $G(w,w_x,w_{xx},\dots)$ are called 
$x-$ and $y-$integrals respectively. For the  detailed discussion of the Darboux integrable equations see \cite{AK}, \cite{ZhiS}, \cite{ZhMHSh} and references therein. 

First, one can look for a  differential-discrete  equation which is  a discretization of a continuous equation (\ref{contEq}). This differential-discrete  equation should also be Darboux integrable.  The notion of Darboux integrable differential-discrete equation was introduced in \cite{HP}. Let us consider an equation
\begin{equation}\label{semidiscr}
t_{1x}=g(t,t_1,t_x),
\end{equation}   
where $t(n,x)$ is a function of a continuous variable $x$, a discrete variable $n$ and $t_1=Dt(n,x)=t(n+1,x)$ ($D$ is the shift operator and $D^kt(n,x)=t(n+k,x)=t_k$, $k\in \mathbb {Z}$). 
Differential-discrete equation (\ref{semidiscr}) is called Darboux integrable if it admits two functions  
$$
I(t,t_x,t_{xx},\dots)\quad \mbox{and} \quad F(\dots, t_{-1},t,t_1,t_2,t_{-2},\dots),
$$ 
depending on a finite number of variables,  such that for all solutions of (\ref{semidiscr}) we have
\begin{equation}
D_xF=0 \quad \mbox{and} \quad DI=I.
\end{equation}
Such Darboux integrable differential-discrete equations and  discrete equations (the definition of Darboux integrable discrete equation is given in the next section)
are actively studied nowadays, see \cite{AS}-\cite{HK}. 
 
It was proposed in \cite{HZhS} to use $x-$ or $y-$ integrals of a continuous equation (\ref{contEq}) to obtain  its discretization. That is, one looks for a differential-discrete  equation that admits a given $x-$ or $y-$integral as its $n-$integral. This approach  allowed to construct many  differential-discrete  equations, see \cite{HZh}-\cite{ZhZh3}. Moreover, the constructed  equations turned out to 
admit also an $x$-integral, that is the equations are Darboux integrable.
Now one can take a constructed differential-discrete  equation and consider its further discretization  using the corresponding $x-$integral.
However, if one tries to find a discrete equation corresponding to a given integral, one obtains  a complicated  functional equation to be solved, see \cite{HZhS} for some examples.  In many cases the discretization is not found.

The Darboux integrability can be also defined in terms of characteristic algebras, which are Lie-Rinehart algebras, see \cite{M, MS, R}. We propose a new approach  which is based on the use of the characteristic algebras.

The paper is organized as follows. In  second section we give necessary definitions  and description of our approach to discretization. In  third section we  give examples of discretization for differential-discrete Darboux integrable equations.

\section{Preliminaries}
In what follows we always assume that $t,t_{\pm 1},t_{\pm 2}, \dots$ and $t_{x},t_{xx}, t_{xxx}\dots$ are independent dynamical variables. Derivatives of variables $t,t_{\pm 1},t_{\pm 2}, \dots$ and shifts of variables $t_x,t_{xx},t_{xxx}, \dots$  are expressed in terms of the dynamical variables using (\ref{semidiscr}).

A criteria for existence of $n-$ and $x-$ integrals of a differential discrete equation (\ref{semidiscr}) can be formulated in terms of the so-called characteristic algebras. 

Let us introduce the criteria for existence of $x-$integral first.
Define an operator
\begin{equation}
Z= t_x\frac{\partial}{\partial t}+t_{1x}\frac{\partial}{\partial t_1}+t_{-1x}\frac{\partial}{\partial t_{-1}}+\dots,
\end{equation}
which corresponds  to the total derivative operator $D_x$ and an operator 
\begin{equation}
W=\frac{\partial}{\partial t_x}. 
\end{equation}
We have that $ZF=0$ and  $WF=0$. Clearly, the function $F$ is also annulled by all possible commutators of these operators. Thus we define  the characteristic $x-$algebra, denoted by $L_x$,  
as a Lie-Rinehart algebra generated by the operators $Z$ and $W$.  The algebra $L_x$ is  considered over the ring of functions depending on a finite 
number of dynamical variables. In general,  all  algebras and linear spaces of operators introduced in this paper are  considered over the ring of functions depending on finite 
number of dynamical variables.
\begin{theorem}\label{th1} \cite{HP}
Equation (\ref{semidiscr}) admits a non-trivial $x-$integral if and only if the corresponding characteristic $x$-algebra $L_x$ is finite dimensional. 
\end{theorem}  
Now we introduce the  criteria  for the existence of the $n-$integral of a differential-discrete equation (\ref{semidiscr}). Following \cite{HP} we define an  operator 

\begin{equation}
Y_0= \frac{\partial}{\partial t_1} 
\end{equation}
and operators
\begin{equation}
Y_k= D^{-k} \frac{\partial}{\partial t_1}  D^{k}, \qquad k=1,2,3, \dots;
\end{equation}
\begin{equation}
X_k= \frac{\partial}{\partial t_{-k}},  \qquad k=1,2,\dots \, .
\end{equation}

\begin{theorem}\label{th2} \cite{HP}
Equation (\ref{semidiscr}) admits a non-trivial $n-$integral if and only if the following conditions are satisfied.\\
1. The linear space generated by the operators $\{Y_k\}_{k=0}^\infty$ has a finite dimension. Let us  denote the dimension by $N$.

\noindent
2. The Lie-Rinehart algebra generated by the operators  $\{Y_k\}_{k=0}^N$ and $\{X_k\}_{k=1}^N$ has a finite dimension.
\end{theorem}
The characteristic $n$-algebra, denoted by  $L_n$, is a Lie-Rinehart algebra generated by the operators  $\{Y_k\}_{k=0}^N$ and $\{X_k\}_{k=1}^N$ from the above theorem.   

Now, we consider a discrete equation case. Assume that a function $u(n,m)$   depends  on two discrete variables $n$ and $m$.
For the function  $u(n,m)$ we have the shift operator  $D$ such that $Du(n,m)=u(n+1,m)=u_1$,  the shift with respect to the first variable, and the shift operator  $\bar D$, $\bar Du(n,m)=u(n,m+1)=u_{\bar 1}$, the shift with respect to the second variable. Note that $D^ku(n,m)=u(n+k,m)=u_k$ and $\bar D^ku(n,m)=u(n,m+k)=u_{\bar k}$, $k\in\mathbb{Z}$.
We study  a discrete equation
\begin{equation}\label{discr}
u_{1\bar1}=f(u,u_1,u_{\bar 1}). 
\end{equation}    
In what follows we always assume that $u,u_{\pm 1},u_{\pm 2}, \dots$ and $u_{\pm\bar 1},u_{\pm\bar 2},\dots$ are independent dynamical variables.  Also $D^k$ shifts of variables $u_{\bar 1},u_{\bar 2},\dots$ and
 $\bar D^k$ shifts of variables $u_1,u_2, \dots$ are expressed in terms of the dynamical variables using (\ref{discr}).

A sequence of functions $\{J_k(u_{-j},\dots,u_r)\}_{k= -\infty}^\infty$, depending on  finite number of dynamical variables $u_{-j},\dots,u_r$, 
is called an $m$-integral for a discrete equation (\ref{discr}) if $\bar D  J_i= J_{i+1}, \quad i\in \mathbb {Z} $.
Note that a shift of an $m-$integral $\{D^pJ_k(u_{-j},\dots,u_r)\}_{k=-\infty}^\infty$, where $p$ is fixed, is also an $m-$integral. 
The notion of an $n$-integral for  a discrete equation (\ref{discr}) is defined in a similar way.
Equation (\ref{discr}) is called Darboux  integrable  if it admits non-trivial  $m$- and $n$-integrals. 
A criteria for existence of $m-$ and $n-$integrals can be formulated in terms of the so-called characteristic algebras. We consider the existence criteria for the  $m-$integral (for the $n-$integral 
the corresponding criteria  is formulated in a similar way).
Following \cite{ZhMHSh}  we define  operators
\begin{equation}
\tilde Y_0= \frac{\partial}{\partial u_{\bar 1}} 
\end{equation}
and
\begin{equation}
\tilde Y_k=\bar D^{-k} \frac{\partial}{\partial u_{\bar 1}} \bar D^{k}, \qquad k=1,2,\dots;
\end{equation}
\begin{equation}
\tilde X_k= \frac{\partial}{\partial u_{-k}},  \qquad k=1,2,\dots \,.
\end{equation}

\begin{theorem}\label{th3} \cite{ZhMHSh}
Equation (\ref{discr}) admits a non-trivial $n-$integral if and only if the following conditions are satisfied\\
1. The linear space generated by the operators $\{\tilde Y_k\}_{k=0}^\infty$ has a finite dimension. Let us  denote the dimension by $\tilde N$.

\noindent
2. The Lie-Rinehart algebra generated by the operators  $\{\tilde Y_k\}_{k=0}^{\tilde N}$ and $\{\tilde X_k\}_{k=1}^{\tilde N}$ has a finite dimension. 
\end{theorem}
The Lie-Rinehart algebra generated by the operators  $\{\tilde Y_k\}_{k=0}^{\tilde N}$ and $\{\tilde X_k\}_{k=1}^{\tilde N}$ from the above theorem is called the characteristic $m$-algebra, denoted by 
$\tilde L_m$. 

Given a Darboux integrable differential-discrete equation (\ref{semidiscr}) we have an $x-$integral $F(t,t_1,\dots,t_j)$. We would like to find a discrete equation  (\ref{discr}) that admits same function $F$ as its $m-$integral. For simplicity we  assume that $\tilde J_k=F$ for all $k$, that is the equality 
\begin{equation}\label{intshift}
\bar D F=F
\end{equation}
holds.  
The equality (\ref{intshift}) in general gives a complicated equation for the function $f$ in (\ref{discr}), namely
\begin{equation}
F(u,u_1,\dots,u_j)=I(u_{\bar 1},u_{1\bar 1},\dots,u_{j\bar 1})=I(u_{\bar 1},f,f(u_1,u_2,f), \dots),
\end{equation}
that is not easy to solve.

To find the discrete equation we propose to use  characteristic algebras. Operators of the characteristic algebra $L_x$ of a given differential-discrete equation and operators of  characteristic algebra $\tilde L_m$ of the corresponding discrete equation annul the same function $F$.  
The operator $\tilde Y_1\in \tilde L_m$ has the form (see \cite{ZhMHSh})
\begin{equation}
\tilde Y_1=\frac{\partial}{\partial u} +\alpha\frac{\partial}{\partial u_1} +\frac{1}{\alpha_{-1}}\frac{\partial}{\partial u_{-1}}+\dots,
\end{equation}
where 
\begin{equation}\label{alpha} 
\alpha=\bar D^{-1} \frac{\partial}{\partial u_{\bar 1}} f.
\end{equation}
We assume that the operator $\tilde Y_1$ can be identified with the operator 
\begin{equation}
[X,Z]=\frac{\partial}{\partial t} +\frac{\partial t_{1x}}{\partial t_x}\frac{\partial}{\partial t_1} + \frac{\partial t_{-1x}}{\partial t_x}\frac{\partial}{\partial t_{-1}} +\dots,
\end{equation}
$[X,Z]\in L_x$ (note that $\displaystyle{\frac{\partial t_{1x}}{\partial t_x}=\left(\frac{\partial t_{-1x}}{\partial t_x}\right)^{-1}}$).
Thus, the coefficient $\displaystyle{\alpha=\bar D^{-1} \frac{\partial}{\partial u_{\bar 1}} f}$ is identified with the coefficient $\displaystyle{\frac{\partial}{\partial t_x} g}$. 
So, if we take function $\displaystyle{\frac{\partial}{\partial t_x} g(t,t_1,t_x)}$  and replace the variables as follows $t=u_{\bar 1}$, $t_1=u_{1 \bar 1}$ and $t_x=A(u,u_1)$ (the function $A$ to be found later) we obtain an equation for $\displaystyle{\frac{\partial}{\partial u_{\bar 1}} f}$
\begin{equation} \label{eqforf}
 \frac{\partial f}{\partial u_{\bar 1}}=\frac{\partial}{\partial t_x} g(t,t_1,t_x)|_{t=u_{\bar 1},\, t_1=u_{1 \bar 1},\, t_x=A(u,u_1)}.
\end{equation} 
The above equation determines  the function $f$ up to some unknown functions of $u, \,u_1$. The unknown functions can be found using  (\ref{intshift}). 

\section{Examples}

In this section we consider the discretization of several differential-discrete Darboux integrable equations.

\begin{example}\label{ex1}
Consider the differential-discrete equation  
\begin{equation}\label{semidiscr1}
t_{1x}=(1+Re^{t+t_1})t_x+\sqrt{R^2e^{2(t+t_1)}+2Re^{t+t_1}}\sqrt{t_x^2-4}
\end{equation}
with the $x$-integral
\begin{equation}
F_1=\sqrt{Re^{2t_1}+2e^{t_1-t}}+\sqrt{Re^{2t_1}+2e^{t_1-t_2}}
\end{equation}
(equation $(3^*b)$ in \cite{HZhS}).
The corresponding discrete equation is given by
\begin{eqnarray}
e^{-u_{1\bar 1}}=e^{-u_{\bar 1}}\left (e^u\sqrt{e^{-u-u_1}+\frac{R}{2}}+e^u\sqrt{\frac{R}{2}} \right)^2  \nonumber\\
+\sqrt{2R}\left (e^u\sqrt{e^{-u-u_1}+\frac{R}{2}}+e^u\sqrt{\frac{R}{2}} \right) \label{discr1}
\end{eqnarray}
Equation (\ref{discr1}) admits the $m$-integral
\begin{equation}
\tilde F_1=\sqrt{Re^{2u_1}+2e^{u_1-u}}+\sqrt{Re^{2u_1}+2e^{u_1-u_2}}
\end{equation}
One can also find an $n$-integral for the given equation
\begin{equation}
\tilde G_1=\frac{e^{-u}+e^{-u_{\bar 1}}}{e^{-u_{\bar 1}}+e^{-u_{\bar 2}}}.
\end{equation}
So, (\ref{discr1}) is Darboux integrable. The derivation of this example is given after Remark \ref{rem1}.
\end{example}

\begin{example} Let us consider the equations of the form
\begin{equation}
t_{1x}=K(t,t_1)t_x
\end{equation}
given in \cite{HZhS} (equations $(2^*b),\,(5^*),\,(6^*)$ ).
Such equations admit an $x$-integral $F(t,t_1)$, where the function $F$ satisfies the equation 
\begin{equation}
F_t+K(t,t_1)F_{t_1}=0.
\end{equation}
For such equations our approach leads to an obvious discrete equation 
\begin{equation}
F(u,u_1)=F(u_{\bar 1},u_{1\bar 1})
\end{equation}
with $m$-integral $F(u,u_1)$. The existence of $n$-integrals for such equations requires further investigation. Some results on a similar classification problems
can be found in \cite{HG}.
\end{example}

\begin{remark}\label{rem1} For  the autonomous equations of the form
\begin{equation}
t_{1x}=t_x+ A(t,t_1)
\end{equation}
given  in \cite{HZhS} (equations $(2^*a),\,(3^*a)$ ).
Our approach leads to a simple discrete equations 
\begin{equation}
u_{1\bar 1}=u_{\bar 1}+u-u_{1},
\end{equation}
in case of equation $(2^*a)$,
and 
\begin{equation}
u_{1\bar 1}=u_{\bar 1}+u_{1}-u,
\end{equation}
in case of equation $(3^*a)$.
\end{remark}

Now let us derive the discrete equation given in the Example \ref{ex1}.
We are looking for a discrete equation (\ref{discr}) which is a discretization of (\ref{semidiscr1}). Since 
\begin{eqnarray} 
\nonumber
\frac{\partial}{\partial t_x} \left( (1+Re^{t+t_1})t_x+\sqrt{R^2e^{2(t+t_1)}+2Re^{t+t_1}}\sqrt{t_x^2-4}\right)\\
=(1+Re^{t+t_1})+\sqrt{R^2e^{2(t+t_1)}+2Re^{t+t_1}}\frac{t_x}{\sqrt{t_x^2-4}},
\end{eqnarray}
we assume that
\begin{equation}\label{eqf1}
\frac{\partial f}{\partial u_{\bar 1}} =1+Re^{u_{\bar 1}+f}+T(u,u_1)\sqrt{R^2e^{2(u_{\bar 1}+f)}+2Re^{{u_{\bar 1}+f}}}
\end{equation} 
where $T$ is a function of $u,u_1$. By solving  (\ref{eqf1}) we get
\begin{equation}
\sqrt{e^{-f-u_{\bar 1}}+\frac{R}{2}}=e^{-u_{\bar 1}}E(u,u_1)+C(u,u_1).
\end{equation}
Since $u_{1\bar1}=f$ we have 
\begin{equation}\label{f1}
\sqrt{e^{-u_{1\bar 1}-u_{\bar 1}}+\frac{R}{2}}=e^{-u_{\bar 1}}E(u,u_1)+C(u,u_1).
\end{equation}
or
\begin{equation}\label{f1.1}
e^{-u_{1\bar 1}}=e^{-u_{\bar 1}}E^2(u,u_1)+ e^{u_{\bar 1}}\left (C^2(u,u_1)-\frac{R}{2}\right) +2E(u,u_1)C(u,u_1).
\end{equation}
To find the functions $E$, $C$ we use equality $\bar D F_1=F_1$. We have
\begin{eqnarray}
  e^{u_{1\bar 1}}\sqrt{e^{-u_{1\bar 1}-u_{\bar 1}}+\frac{R}{2}}+e^{u_{1\bar 1}}\sqrt{e^{-u_{1\bar 1}-u_{2\bar 1}}+\frac{R}{2}} \nonumber\\
=e^{u_{1}}\sqrt{e^{-u_{1}-u}+\frac{R}{2}}+e^{u_{1}}\sqrt{e^{-u_{1}-u_{2}}+\frac{R}{2}}. \label{F1}
\end{eqnarray}
Using  (\ref{f1}) we can write 
\begin{equation}
\sqrt{e^{-u_{2\bar 1}-u_{1\bar 1}}+\frac{R}{2}}=e^{-u_{1\bar 1}}E(u_1,u_2)+C(u_1,u_2)
\end{equation}
and rewrite  (\ref{F1}) as
\begin{eqnarray} 
\nonumber
e^{u_{1\bar 1}}(e^{-u_{\bar 1}}E(u,u_1)+C(u,u_1))+e^{u_{1\bar 1}}(e^{-u_{1\bar 1}}E(u_1,u_2)+C(u_1,u_2))\\
=e^{u_{1}}\sqrt{e^{-u_{1}-u}+\frac{R}{2}}+e^{u_{1}}\sqrt{e^{-u_{1}-u_{2}}+\frac{R}{2}}. \label{F1.1}
\end{eqnarray}
By differentiating  (\ref{F1.1}) with respect to  $u_{\bar 1}$ we get 
\begin{equation}
\frac{\partial u_{1\bar 1}}{\partial u_{\bar 1}}=\frac{e^{-u_{\bar 1}}E(u,u_1)}{e^{-u_{\bar 1}}E(u,u_1)+C(u,u_1)+C(u_1,u_2)},
\end{equation}
which implies that
\begin{equation}
u_{1\bar 1}=-\ln(e^{-u_{\bar 1}}E(u,u_1)+C(u,u_1)+C(u_1,u_2)) +\hat K(u,u_1)
\end{equation}
or
\begin{equation}\label{e1}
e^{-u_{1\bar 1}}=K(u,u_1)(e^{-u_{\bar 1}}E(u,u_1)+C(u,u_1)+C(u_1,u_2)). 
\end{equation}
Now comparing  (\ref{e1}) with (\ref{f1.1}) we get $C(u,u_1)=\pm \sqrt{\frac{R}{2}}$ and $K(u,u_1)=E(u,u_1)$.
Thus,
\begin{equation}\label{f1.2}
e^{-u_{1\bar 1}}=E^2(u,u_1)e^{-u_{\bar 1}} \pm\sqrt{2R}E(u,u_1). 
\end{equation}
By substituting the above expression for $e^{-u_{1\bar 1}}$ into  (\ref{F1.1}) and differentiating (\ref{F1.1}) with respect to  $u_{2}$ we get 
\begin{equation}
\frac{\partial E(u_1,u_2)}{\partial u_{ 2}}=\frac{\partial}{\partial u_{ 2}}\left(e^{u_1} \sqrt{e^{-u_1-u_2} +\frac{R}{2}}\right),
\end{equation}
that is
\begin{equation}
 E(u_1,u_2)=e^{u_1} \sqrt{e^{-u_1-u_2} +\frac{R}{2}} +M(u_1)
\end{equation}
So, we rewrite  (\ref{f1.2}) as
\begin{equation}\label{f1.3}
e^{-u_{1\bar 1}}=e^{-u_{\bar 1}}(e^{u} \sqrt{e^{-u-u_1} +\frac{R}{2}} +M(u))^2 \pm\sqrt{2R}e^{u} \sqrt{e^{-u-u_1} +\frac{R}{2}} +M(u). 
\end{equation} 
We use  (\ref{F1}) and have $M(u)=e^u\sqrt{\frac{R}{2}}$. Thus, we obtain (\ref{discr1}).


\begin{thebibliography}{100}


\bibitem {AK}  Anderson I M and  Kamran N 1997  The variational bicomplex for hyperbolic second-order scalar partial differential equations in the plane {\it  Duke Math. J.} \textbf {87} 265-319

\bibitem{AS} Adler V E and Startsev S Ya  1999 On discrete analogues of the Liouville equation {\it Theoret. and Math. Phys.} \textbf {121} 1484-1495

\bibitem{F} Ferapontov E V, Habibullin I T, Kuznetsova M N and Novikov V S 2020 On a class of 2D integrable lattice equations {\it J. Math. Phys.} \textbf {61} 073505

\bibitem{GY1} Garifullin R N and Yamilov R I  2020 Modified series of integrable discrete equations on a quadratic lattice with a nonstandard symmetry structure {\it Theoret. and Math. Phys.}
 \textbf{205} 1264-1278 

\bibitem {GY2} Garifullin R N and Yamilov R I  2019 On series of Darboux integrable discrete equations on square lattice {\it Ufa Math. J.} \textbf {11} 99-108

\bibitem{HK} Habibullin I T and Kuznetsova M N 2021  An algebraic criterion of the Darboux integrability of differential-difference equations and systems {\it J. Phys. A} \textbf {54}  505201

\bibitem{HZhP2} Habibullin I T,  Zheltukhina N  and Pekcan A 2008 On the classification of Darboux integrable chains  {\it J. Math. Phys.} \textbf {49} 102702

\bibitem{HZhS} Habibullin I T,  Zheltukhina N and Sakieva A 2011 Discretization of hyperbolic type Darboux integrable equations preserving integrability {\it J. Math. Phys.} \textbf {52}  093507

\bibitem{HZh} Habibullin I T and Zheltukhina N 2016  Discretization of Liouville type nonautonomous equations {\it J. Nonlinear Math. Phys.} \textbf {23} 620-642

\bibitem{HP} Khabibullin I T and Pekcan A   2007 Characteristic Lie algebra and the classification of semi-discrete models {\it Theoret. and Math. Phys.} \textbf {151} 781-790

\bibitem{M} Millionshchikov D  2018 Lie algebras of slow growth and Klein-Gordon PDE {\it Algebr. Represent. Theory} \textbf{21} 1037-1069

\bibitem{MS}  Millionshchikov D V and Smirnov S V 2021  Characteristic algebras and integrable exponential systems  {\it Ufa Math. J.} \textbf {13}  41-69

\bibitem{R} Rinehart G  1963 Differential forms for general commutative algebras {\it Trans. Amer. Math. Soc.} \textbf{108} 195-222

\bibitem{ZhZh1} Zheltukhin K and Zheltukhina N 2021 On discretization of Darboux integrable systems admitting second-order integrals {\it Ufa Math. J.} \textbf {13}  170-186

\bibitem{ZhZh2} Zheltukhin K and Zheltukhina N 2020  On the discretization of Darboux integrable systems {\it J. Nonlinear Math. Phys.} \textbf {27} 616-632

\bibitem{ZhZh3}  Zheltukhin K and Zheltukhina N 2018 On the discretization of Laine equations {\it J. Nonlinear Math. Phys.} \textbf {25} 166-177

\bibitem{ZhiS}  Zhiber A V and  Sokolov V V  2001 Exactly integrable hyperbolic equations of Liouville type {\it Russian Math. Surveys} \textbf {56} 61-101

\bibitem{ZhMHSh} Zhiber A V,  Murtazina R D,  Habibullin I T  and   Shabat A B  2012 Characteristic Lie rings and integrable models in mathematical physics {\it Ufa Math. J.} \textbf{4} 17-85

\bibitem{HG} Habibullin I T and Gudkova E V  2011 An algebraic method for classifying S-integrable discrete models {\it Theoret. and Math. Phys.} \textbf {167} 407-419
 
\end{thebibliography}
\end{document}